\documentclass[10pt]{article}

\usepackage{amsmath}
\usepackage{color}
\usepackage[colorlinks,citecolor=blue,urlcolor=blue,linkcolor=blue]{hyperref}
\usepackage[utf8]{inputenc}
\usepackage{graphicx}
\usepackage{float}
\usepackage{bm}
\usepackage{enumerate}
\usepackage{authblk}
\usepackage{notes2bib}
\usepackage{braket}

\numberwithin{equation}{section}

\newcommand{\Hb}{\overline{\mathrm{H}}}
\newcommand{\gb}{\overline{g}}
\newcommand{\pb}{\overline{p}}
\newcommand{\eb}{\overline{e}}
\newcommand{\D}{\mathrm{d}}
\newcommand{\micro}{$\mu$}

\begin{document}

\title{Prospects for studies of the free fall and gravitational quantum states of antimatter}
\date{}

\author[1]{G.~Dufour}
\author[2]{D.B.~Cassidy}
\author[3]{P.~Crivelli}
\author[4]{P.~Debu}
\author[1]{A.~Lambrecht}
\author[5]{V.V.~Nesvizhevsky}
\author[1]{S.~Reynaud}
\author[6]{A.Yu.~Voronin}
\author[2]{T.E.~Wall}

\affil[1]{LKB - CNRS, ENS, UPMC, Paris, France}
\affil[2]{University College, London, UK}
\affil[3]{ETH, Zurich, Switzerland}
\affil[4]{CEA-IRFU, Saclay, France }
\affil[5]{ILL, Grenoble, France }
\affil[6]{Lebedev Institute, Moscow, Russia }

\maketitle

\begin{abstract}

Different experiments are ongoing to measure the effect of gravity
on cold neutral antimatter atoms such as positronium, muonium and
antihydrogen. Among those, the project GBAR in CERN aims to measure
precisely the gravitational fall of ultracold antihydrogen atoms. In
the ultracold regime, the interaction of antihydrogen atoms with a
surface is governed by the phenomenon of quantum reflection which
results in bouncing of antihydrogen atoms on matter surfaces. This
allows the application of a filtering scheme to increase the
precision of the free fall measurement. In the ultimate limit of
smallest vertical velocities, antihydrogen atoms are settled in
gravitational quantum states in close analogy to ultracold neutrons
(UCNs). Positronium is another neutral system involving antimatter
for which free fall under gravity is currently being investigated at
UCL. Building on the experimental techniques under development for
the free fall measurement, gravitational quantum states could also
be observed in positronium. In this contribution, we review the
status of the ongoing experiments and discuss the prospects of
observing gravitational quantum states of antimatter and their
implications.
\end{abstract}

\newpage

\emph{This work reviews contributions made at the GRANIT 2014
workshop on prospects for the observation of the free fall and
gravitational quantum states of antimatter.}

\section*{Introduction}

At present, together with the dark matter problem, one of the most
tantalizing open questions in physics is the baryon-antibaryon
asymmetry, i.e. why are we living in a matter-dominated Universe?
Where did all the antimatter go? Different theoretical and
experimental efforts trying to address this question are ongoing,
including activities focusing on the gravitational behavior of
antimatter
\cite{TheALPHACollaboration2013,AEGISCollaboration2012,Perez2012,Kirch2014,Cassidy2014}.
No compelling theoretical argument seems to support that a
difference between the gravitational behavior of matter and
antimatter should be expected \cite{Nieto1991}, although some
attempts have been made to show the contrary
\cite{Scherk1979,Chardin1992,Chardin1993,Chardin1997,Kostelecky2011}.
Moreover observations and experiments have been interpreted as
evidence against the existence of ``antigravity'' type forces
\cite{Good1961,Pakvasa1989,Apostolakis1999,Gabrielse1999}. However
those could be argued to be model dependent and therefore a simple
free fall measurement is preferable. This justifies ongoing
experimental efforts in that direction. A first attempt in this
direction was made recently by the ALPHA collaboration
\cite{TheALPHACollaboration2013} that bounds the ratio of
gravitational mass to inertial mass of antihydrogen between $-$65
and 110.

The idea of directly measuring the gravitational force acting on
antiparticles in the Earth's field goes back many decades, from the
work of Witteborn and Fairbank, attempting to measure the
acceleration of gravity for electrons and, eventually, positrons
\cite{Witteborn1967} to the PS200 experiment at CERN in the 1980's
that included measurements on antiprotons \cite{Holzscheiter2014}.
Such measurements are extremely difficult because measuring the
force of gravity on a charged particle requires a physically
unrealistic (it would seem) elimination of stray electromagnetic
fields. The obvious solution to this problem is to use neutral
antimatter particles. However, at the present time it is not
technically feasible to do so; antineutrons cannot be produced in a
controllable manner and antineutrinos are similarly elusive to
experimenters. One may instead consider using composite systems that
are electrically neutral, in which case it is only necessary to
contend with dipole moments. Only a few systems that are composed
of, or contain some fraction of, antimatter are available for
scientific study. These are antihydrogen, muonium and positronium,
which have all been suggested as possible candidates for gravity
measurements \bibnote{See for example articles 4-6 in Antimatter and
Gravity, \emph{International Journal of Modern Physics: Conference
Series}, 30, January 2014}.

Selecting between different experimental methods, one should aim at
precision experiments as they are much more strongly motivated
theoretically. Among these, the method of quantum gravitational
spectroscopy stands out by its remarkable statistical sensitivity
and its cleanness from a systematic point of view.

Gravitational quantum states are solutions of the Schr\"odinger
equation in a gravity field above a surface. They are characterized
by the following energy ($E_n$) and spatial scale ($H_n$) :
\begin{align}
&E_n=\varepsilon_0 \lambda_n,\qquad \quad
\varepsilon_0=\sqrt[3]{\frac{\hbar^2M^2g^2}{2m}}~,
\label{eq:En}\\
&H_n=E_n/Mg ~,\label{eq:Hn} \\
&\mathop{\rm Ai}(-\lambda_n)=0,\qquad \lambda_n\approx\{2.34,4.09,
5.52, 6.79, 7.94, 9.02, 10.04...\}~.
\end{align}
Here $M$ is the gravitational mass of the particle, $m$ is its
inertial mass (we distinguish between $M$ and $m$ in view of
discussing EP tests), $g$ is the gravitational field
\emph{intensity} near the Earth surface, $\gb=Mg/m$ is the
acceleration of the particle in that field and $\mathop{\rm Ai}(x)$
is the Airy function \cite{Abramowitz1972,Landau1965}. For neutrons
and antihydrogen atoms, the height of the lowest gravitational level
is 13.7 \micro m. For positronium, whose mass is approximately 1000
times smaller, it extends over 1.3 mm. The frequency of transitions
between first and second quantum states equals 254 Hz for neutrons
and antihydrogen, and 26 Hz for positronium. The corresponding
characteristic times needed to form quantum states are 0.5 ms and 5
ms respectively.

Quantum gravitational states were observed for the first time with
neutrons by measuring their transmission through a slit made of a
mirror and an absorber in the GRANIT experiment
\cite{Nesvizhevsky2002nature}. If the distance between the mirror
and the absorber (which is a rough surface used as a scatterer to
mix the velocity components) is much higher than the turning point
for the corresponding gravitational quantum state, the neutrons pass
through the slit without significant losses. As the slit size
decreases the absorber starts approaching the size of the neutron
wave function and the probability of neutron loss increases. If the
slit size is smaller than the characteristic size of the neutron
wave function in the lowest quantum state, the slit is not
transparent for neutrons as was demonstrated experimentally.

Here we analyze in detail several experiments which will study the
free fall of anti-atoms and argue that the observation of
gravitational quantum states of antimatter is feasible. In section
\ref{sec:gbar} we describe the forthcoming $\Hb$ experiment GBAR. We
explain in section \ref{sec:qrefl} the quantum reflection mechanism
which allows the formation of gravitational quantum states of $\Hb$
above material surfaces. In section \ref{sec:shaper} we show how the
filtering scheme of the GRANIT experiment could be implemented in
GBAR and in section \ref{sec:gravstateshbar} we describe a possible
spectroscopy of gravitational quantum states of $\Hb$. Section
\ref{sec:gravfallps} reviews the status of positronium free fall
experiment at UCL and section \ref{sec:gravstatesps} explores the
possibility of observing gravitational quantum states of
positronium.

\section{GBAR status report}\label{sec:gbar}

GBAR is an experiment in preparation at CERN. Its goal is to measure
the gravitational acceleration ($\gb=Mg/m$) imparted to freely
falling antihydrogen atoms, in order to perform a direct
experimental test of the Weak Equivalence Principle with antimatter.
The objective is to reach a relative precision on $\gb$ of $1\%$ in
a first stage, with the perspective to reach a much higher precision
using quantum gravitational states in a second stage, as is
described in section \ref{sec:gravstateshbar}.

The principle of the experiment is described in detail in
\cite{Chardin2011} and is briefly recalled here. It is based on an
idea proposed in \cite{Walz2004}. Antihydrogen ions $\Hb^+$  are
produced, trapped and sympathetically cooled to around 10 \micro K.
The excess positron is detached by a laser pulse, which gives the
start signal for the free fall of the ultracold antihydrogen atom
$\Hb$. The  $\Hb$ subsequent annihilation on a plate is detected and
provides the information to measure  $\gb$. The choice of producing
$\Hb^+$ ions to get ultracold antihydrogen atom is the specificity
of the GBAR experiment. It is very costly in statistics, but makes
the cooling to \micro K temperatures a realistic aim.

We report in this section on three recent progresses in the
preparation of the experiment: estimations of the $\Hb^+$ production
cross sections, accumulation of positrons, and cooling of the
$\Hb^+$ ions.

\subsection{Production cross sections of $\Hb^+$ ions}

The $\Hb$ production proceeds in two steps: $\pb + \mathrm{Ps} \to
\Hb + \mathrm{e}^-$ (1) followed by $\Hb+\mathrm{Ps} \to \Hb^+ +
\mathrm{e}^-$ (2). The Ps symbol stands for positronium. The
cross-sections of these reactions are not well known and are very
low.  The matter counterpart of the first one has been measured. It
is around $10^{-15}$~cm$^2$ (10$^9$~barn) for tens of keV protons
\cite{Merrison1997}. The second one is estimated to be around
$10^{-16}$~cm$^2$ (10$^8$~barn) \cite{Walters2007}.

New calculations of these reactions have been performed in which the
first excitations levels for the Ps (up to $n=3$)  and the $\Hb$ (up
to $n=5$ ) have been considered. The results suggest that the
production of $\Hb^+$ can be efficiently enhanced by using either a
fraction of Ps(2p) and a 2 keV antiproton beam or a fraction Ps(3d)
and antiprotons with kinetic energy below 1 keV \cite{Comini2013}.
The product of the cross sections of reactions (1) and (2) reaches
values around  $10^{-29}$~cm$^4$ (10$^{19}$~barn$^2$) for an
optimized fraction of excited Ps. Simulations are underway to
estimate the effective gain with a realistic experimental setup.

This shows that very low energy antiprotons are needed. The
Extremely Low Energy Antiproton (ELENA) ring which is in
construction at CERN and which will complement the Antiproton
Decelerator (AD), will provide 75 ns rms bunches of $5\times 10^{6}$
100 keV antiprotons every 100 s.  Those have to be further
decelerated and cooled to match the GBAR requirements. The
decelerator is under construction at CSNSM (``Centre de Sciences
Nucl\'eaires et de Sciences de la Mati\`ere'') in Orsay, France.

\subsection{Positron accumulation}
In addition to a high flux of low energy antiprotons, the production
of $\Hb^+$ via reactions (1) and (2) require to form a dense cloud
of positronium. It has been shown that Ps can be efficiently
produced by dumping few keV positrons on mesoporous silica films.
Yields of 30 to 40 $\%$ depending on the incident positron energy
(few keV) have been measured \cite{Liszkay2008, Cassidy2011}. The
accumulation of a very large number of positrons, around $2 \times
10^{10}$, between two ejections of antiprotons from ELENA is thus
necessary to produce a dense enough positronium cloud.

A demonstration facility for the production and accumulation of
positrons is currently running at CEA/Saclay. It consists of a low
energy electron linear accelerator (LINAC), a high field
Penning-Malmberg trap from the Atomic Physics Laboratory in RIKEN,
Japan, and a dedicated beam line for further studies of
positron-positronium conversion and for applications in material
science. In addition, a laser system is now being built at LKB
(``Laboratoire Kastler-Brossel'') in Paris to test the excitation of
the positronium which will be formed downstream of the trap.

The LINAC produces a 4.3 MeV electron bunched beam. The bunch length
is 200 \micro s, and the LINAC runs at 200 Hz, producing a mean
current of 120 \micro A. Electrons are sent onto a tungsten mesh
moderator. A flux of typically $3\times {{10}^{6}}$ slow (few 100
eV) positrons per second is driven towards the Penning trap through
a vacuum tube equipped with solenoid coils producing a 80 mT field.
They are accelerated to around 1 keV to enter the high magnetic
field (5 T) region. They reach the Penning trap which is made of 23
cylindrical electrodes, surrounded by 4 additional long electrodes
to control the admission and the trapping of incident particles.
Positrons make a round trip in less than 100 ns. In order to trap
them, it is necessary to compress the 200 \micro s  bunch. This is
done by applying a varying voltage (20 to 150 V) when extracting the
slow positron from the moderator. In this way, it is possible to
close the entrance of the trap before the bunch escapes. This method
allows to trap one single bunch.

\begin{figure}
\centering
 \includegraphics[width=0.80\textwidth]{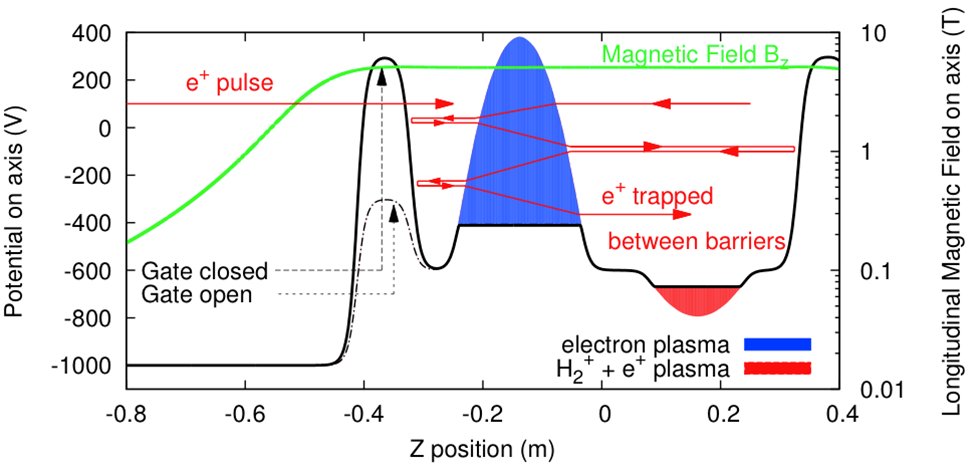}
\caption{\label{fig:trapping} The positron trapping mechanism (from
\cite{Grandemange2013}). The horizontal axis is the position along
the trap axis. The left vertical axis is the voltage seen by
particles. The  magnetic field strength is shown by the green curve,
with the scale on the right vertical axis. In blue, the electron
potential well filled with electrons is drawn , reducing the
apparent voltage shown by black curve with the value on the left
vertical axis. In red is shown the positron potential well. When a
positron bunch arrives, the entrance electrode voltage is low
(dashed dotted line). It is then increased and positrons go back and
forth (it is depicted by red arrows) between this gate and the
downstream part of the trap. They pass many times through the
electron plasma, and are eventually slowed down and fall into their
well. The presence of residual $\mathrm{H} _{\mathrm{2}}
^{\mathrm{+}}$ helps the final catching of positrons.}
\end{figure}

In order to accumulate a large number of bunches, positrons have to
be slowed down and stored in a dedicated potential well formed by a
subset of electrodes of the trap (see Fig. \ref{fig:trapping})
before the next bunch arrives. Positrons are cooled by passing
through a preloaded electron plasma in another dedicated potential
well. This method has been set up and demonstrated in
\cite{Oshima2004} with a continuous positron beam issued from a
$^{22}$Na source. With such a beam, it is not possible to close the
entrance gate, and positrons must be slowed down in one step. This
was done with a remoderator downstream of the trap. An efficiency of
1$\%$ was obtained. With a bunched positron beam, the remoderator is
not necessary, and a much higher efficiency is expected.

The cooling time fixes the maximal LINAC frequency, and depends on
the density of the electron plasma. With  ${{10}^{17}}$ e$^-$/m$^3$,
simulations show that the cooling time is around 3 ms.

\begin{figure}
\centering
 \includegraphics[width=0.80\textwidth]{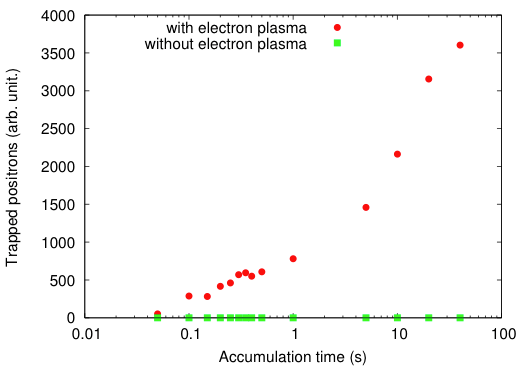}
\caption{\label{fig:acc} The accumulation of positrons (from \cite{Grandemange2013}).}
\end{figure}

The principle of this accumulation scheme has now been successfully
demonstrated at Saclay. The details of the experimental setup used
during accumulation are given in \cite{Grandemange2013}. The result
of a successful set of accumulation trials is shown in Fig.
\ref{fig:acc}.
Given the characteristics of the demonstrator facility at Saclay, a
realistic objective is now to accumulate around 10$^8$ positrons in
the trap within 2 minutes.

\subsection{Cooling of the $\Hb^+$ ions}

Recent progresses have been made for the design of the cooling of
$\Hb^+$ions. The cooling proceeds in two steps: Doppler cooling at
the mK level, Raman side band cooling to reach 10 \micro K.

In the first step, ions are captured in a linear Paul trap inside
which Be$^+$ ions are preloaded and laser cooled. In the original
scheme of GBAR, it was assumed that the $\Hb^+$ ions would be cooled
by Coulomb interactions with the Be$^+$ ions (``sympathetic
cooling'').  Simulations show that this process is very slow. This
is due to the large mass ratio between the ions.  As a consequence,
one cannot reach the mK level in a short enough time to avoid the
destruction of the $\Hb^+$: the laser cooling of Be$^+$ induces the
photodetachment of the excess positron in a fraction of a second.
However, the simulations show also that the addition of a third
species of ions of intermediate mass, namely HD$^+$ ions, make the
process efficient enough \cite{Hilico2014}. Starting with 1800
Be$^+$ and 200 HD$^+$ ions, cooling times of ms are achievable.

Second step: to reach the 10 \micro K level necessary for the free
fall experiment, a Be$^+$/$\Hb^+$ ion pair must be transferred to a
precision trap to undergo a ground state Raman side band cooling.
Calculations show that one may achieve the desired cooling in less
than a second. This is  shown in \cite{Hilico2014} and references
therein. This method will be tested with matter ions (Ca$^+$ /Be$^+$
, H$_2 ^+$ /Be$^+$) before being implemented for the GBAR
experiment. Traps are being mounted at LKB in Paris and at Mainz
University.

Since the uncertainty on the measurement of $\gb$ is fully dominated
by the initial velocity dispersion due to both the vertical velocity
after cooling and the recoil due to the positron photodetachment,
the implementation of a vertical velocity selector will allow a
drastic gain in the statistics needed to reach the 1$\%$ precision
on $\gb$ as is described in section \ref{sec:shaper}.

\section{Quantum reflection of antihydrogen on material surfaces}\label{sec:qrefl}

In the ultracold regime, the interaction of antihydrogen ($\Hb$)
atoms with a surface is governed by the phenomenon of quantum
reflection. Although the atoms are strongly attracted to the
surface, the atomic wave function can be partly reflected on the
steep atom-surface potential, leading to a non-zero probability of
classically forbidden reflection. This effect is relevant to
experiments such as GBAR where ultracold  $\Hb$ atoms are detected
by annihilation on a plate (see section \ref{sec:gbar}).

A single atom placed in vacuum near a material surface experiences
an attractive Casimir-Polder (CP) force \cite{Casimir1946,
Casimir1948}. This force is a manifestation of the electromagnetic
quantum fluctuations which are coupled to the atomic dipole. Quantum
reflection occurs if an atom impinges with low velocity on such a
rapidly varying potential \cite{Friedrich2002}. We will give a more
explicit condition later on.

In this section we first describe how the CP potential is calculated
for realistic experimental conditions. We then go on to compute the
scattering amplitudes of an  atom on this potential. We show that
quantum reflection can be understood as a deviation from the
semiclassical approximation. Finally we describe materials from
which quantum reflection is enhanced and above which gravitationally
bound states of $\Hb$ could be observed.

\subsection{Calculation of the Casimir-Polder potential}

We use the scattering approach to Casimir forces
\cite{Lambrecht2006} to give a realistic estimation of the
atom-surface interaction energy. In this approach the interacting
objects are described by reflection matrices for the electromagnetic
field. Reflection on a plane is described by Fresnel coefficients,
while reflection on the atom is treated in the dipolar approximation
and depends on the dynamic polarizability \cite{Messina2009}. This
allows an evaluation of the CP potential for any material when its
optical properties are known. Those used here are detailed in
\cite{Dufour2013qrefl}. Note that since the typical length scale for
quantum reflection ($\sim$100 nm) is below the thermal wavelength at
300 K ($\sim$1 \micro m), we carried out all calculations at null
temperature.

\begin{figure}
\centering
\includegraphics[width=0.80\textwidth]{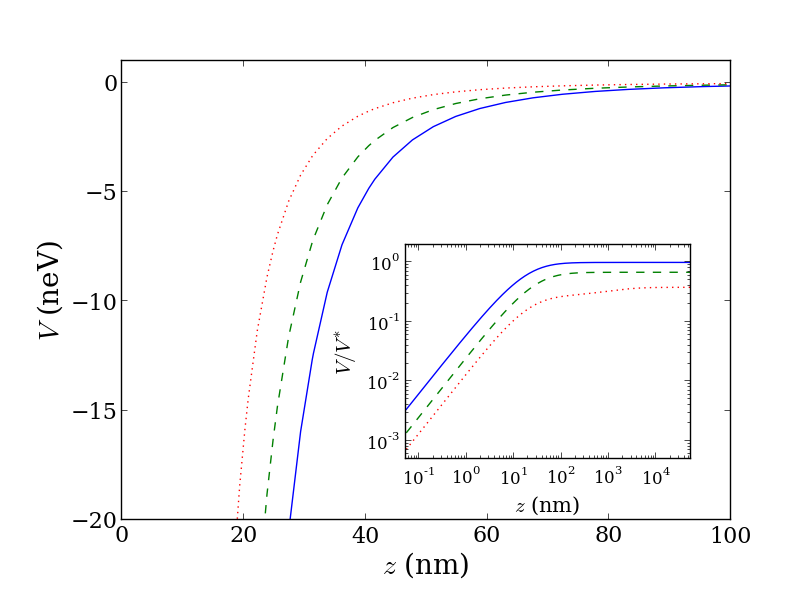}
\caption{\label{fig:CPpotential} Casimir-Polder (CP) potential for
antihydrogen  in the vicinity of a material bulk; from top to
bottom, perfect conductor (full line), silicon (dashed line), silica
(dotted line); (inset:ratio $V/V^*$ to the retarded potential $V^*$
for a perfectly conducting mirror, see text).}
\end{figure}

The CP potential for $\Hb$ at a distance $x$ of a perfectly
conducting plane and thick silicon and silica slabs are presented in
Fig. \ref{fig:CPpotential}. For a perfectly conducting mirror in the
long-distance regime we recover the historic result of Casimir and
Polder \cite{Casimir1946,Casimir1948}:
\begin{equation}
V(x) \underset{x \gg \lambda}{\approx} V^*(x) = - \frac{3 \hbar c}{8
\pi x^4} \frac{\alpha(0)}{4\pi\epsilon_0}~,
\end{equation}
where $\alpha(0)$ is the static polarizability of the atom.

For real mirrors, the potential is reduced but it shows the same
power law dependence in the van der Waals (short distance) and
retarded (long distance) regimes:
\begin{equation}
V(x) \underset{x \ll \lambda}{\approx} - \frac{C_3}{x^3} ~, \qquad
\qquad V(x) \underset{x \gg \lambda}{\approx} -\frac{C_4}{x^4}~,
\end{equation}
where $\lambda$ is a typical wavelength associated with the optical
response of atom and plane.

\subsection{Scattering on the Casimir-Polder potential}

We now solve the Schr\"odinger equation for an atom of energy $E>0$
scattering on the CP potential $V(x)$ :
\begin{equation}
\frac{\textrm{d}^2  }{\textrm{d} x^2} \psi(x) +
\frac{p(x)^2}{\hbar^2} \psi(x) = 0 ~,
\end{equation}
with $p(x) = \sqrt{2 m (E - V(x))}$ the classical momentum. We write
the exact wave function as a sum of counter-propagating WKB waves
whose coefficients are allowed to vary:
\begin{equation}\label{eq:wkbbasis}
\psi(x) = \frac{c_{in}(x)}{\sqrt{p(x)}}
\exp \left(-\frac{i}{\hbar} \int^x p(x ') \D x ' \right)
+ \frac{c_{out}(x)}{\sqrt{p(x)}} \exp \left(\frac{i}{\hbar}
\int^x p(x ') \D x ' \right)~.
\end{equation}

Upon insertion in the Schr\"odinger equation we obtain coupled first
order equations for the coefficients $c_{in}(x),c_{out}(x)$
\cite{Berry1972}. The annihilation of $\Hb$ on the material surface
translates as a fully absorbing boundary condition on the surface:
$c_{out}(x=0)=0$. This is in contrast with matter atoms, for which
more complicated surface physics is involved in the boundary
condition. Close to the surface, the energy becomes negligible
compared with the potential, which takes the van der Waals form and
$c_{in}(x),c_{out}(x)$ can be solved for analytically
\cite{Dufour2013qrefl}. The equations are then integrated
numerically until $c_{in}(x),c_{out}(x)$ become constants.

\begin{figure}
\centering
\includegraphics[width=0.80\textwidth]{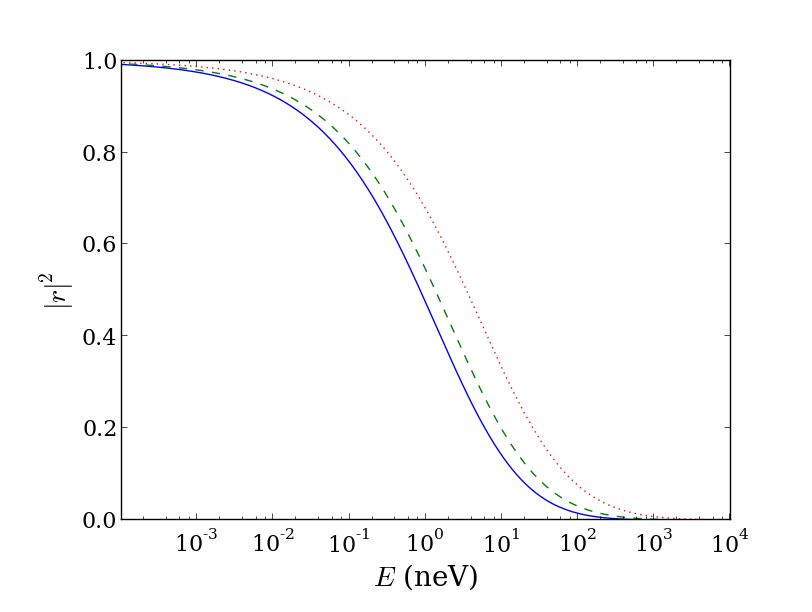}
\caption{\label{fig:reflectivity}  Quantum reflection probability
$|r|^2$ as a function of the energy for antihydrogen atoms on bulk
mirrors; from bottom to top, perfect conductor (full line), silicon
(dashed line), silica (dotted line).}
\end{figure}

The reflection probability
$|r|^2=\underset{x\to\infty}{\lim}|c_{out}(x)/c_{in}(x)|^2$ is
plotted against the energy $E$ in Fig. \ref{fig:reflectivity} for
various semi-infinite media. Note that the quantum reflection
probability is larger for materials with a weaker CP interaction,
such as silica.

\subsection{The badlands function}

To understand this surprising result we look more closely at what
distinguishes the exact solution of the Schr\"odinger equation from
the reflectionless WKB approximation. If $c_{in},c_{out}$ are no
longer allowed to vary, one can show that the wave function
\eqref{eq:wkbbasis} obeys a modified Schr\"odinger equation where
$p(x)^2$ is replaced by $\tilde p(x)^2 = p(x)^2(1+Q(x))$
\cite{Berry1972}. $Q(x)$ is known as the badlands function since the
WKB approximation is not valid in regions where it is
non-negligible:
\begin{equation}
Q(x) = \frac{\hbar^2}{2 p(x)^2} \left(\frac{p''(x)}{p(x)}
- \frac{3}{2} \frac{p'(x)^2}{p(x)^2} \right)~.
\end{equation}

For the CP potential, the badlands function exhibits a peak in the
region where $|V(x)|=E$ but goes to zero both as $x\to\infty$ (where
the potential cancels) and as $x\to0$ (where the classical momentum
diverges).

\begin{figure}
\centering
\includegraphics[width=0.80\textwidth]{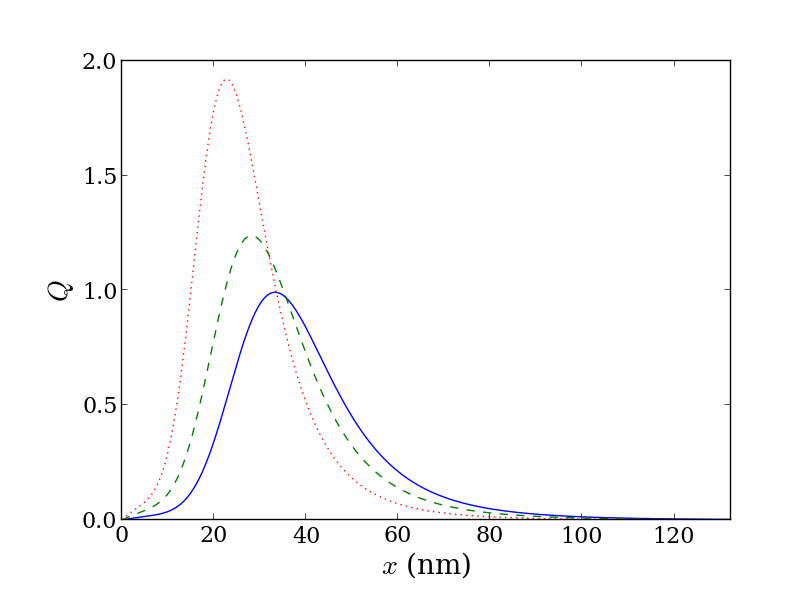}
\caption{\label{fig:badlands} The badlands function $Q(x)$ for an
antihydrogen atom with energy $E=10$~neV; from bottom to top,
perfect conductor (full line), silicon (dashed line), silica (dotted
line).}
\end{figure}

As the energy is decreased, the semiclassical approximation breaks
down and the  badlands function's peak becomes larger. But for a
given energy, the peak is larger and closer to the surface when the
potential is weak, as shown in Fig. \ref{fig:badlands}. The
difference between exact and WKB solutions is larger in weaker CP
potentials, leading to enhanced quantum reflection.

\subsection{Enhancing quantum reflection}

Quantum reflection first appears as a bias in the context of the
GBAR experiment, since it tends to exclude low energy atoms from the
statistics. However this phenomenon opens perspectives for the
storage and guiding of antimatter with material walls. With this in
mind we consider materials which couple weakly to the
electromagnetic field and are therefore good mirrors for atoms, as
we have seen in the previous paragraph.

A simple strategy is to remove matter from the reflective medium, by
using thin slabs or porous materials for example. Our versatile
approach allowed us to compute the CP interaction near thin slabs,
an undoped graphene sheet \cite{Dufour2013qrefl} and nanoporous
materials \cite{Dufour2013porous}. The latter consist in a  solid
matrix which forms an array of nanometric pores. Aerogels, which are
obtained by supercritically drying a silica gel, are a well known
example. We also consider porous silicon and powders of diamond
nanoparticles formed by explosive shock.

From a distance larger than the typical pore size, such materials
can be modeled as homogeneous effective media with properties
averaged between that of vacuum and of the solid matrix. In
consequence their effective dielectric constant is extremely low, as
a result of which quantum reflection is exceptionally efficient. In
table \ref{tab:lifetimes} we show the lifetime of an antihydrogen
atom in the first gravitationally bound state above a surface (see
section \ref{sec:gravstateshbar} for more details).

\begin{table}
  \centering
  \begin{tabular}{c|c}
    Surface (porosity)  & Lifetime (s)\\
  \hline\hline
    perfect conductor  & 0.11\\ \hline
    bulk silicon  &  0.14\\ \hline
    bulk silica  &  0.22\\ \hline
    nano-diamond powder (95\%) & 0.89\\ \hline
    porous silicon (95\%) &  0.94\\ \hline
    silica aerogel (98\%) &  4.6
  \end{tabular}

\caption{\label{tab:lifetimes} The lifetime of the first
gravitationally bound state of antihydrogen above various surfaces.}
\end{table}

Note that this approach does not take into account the possible
presence of stray charges on the surface, a question that would have
to be addressed to observe the predicted reflection probabilities.
Moreover, the effective medium approximation is applicable only for
low atom velocities, such that the atom is reflected far enough from
the surface. With these caveats, nanoporous materials are an
outstanding candidate for the manipulation and study of antihydrogen
and its gravitationally bound states, with lifetimes above the
second.

\section{Shaping of vertical velocity components of antihydrogen atoms for GBAR}
\label{sec:shaper}

The main source of uncertainty on the determination of $\gb$ in the
GBAR experiment is the width of the vertical velocity distribution
of the atom at the beginning of the free fall. This spread in
velocities is due to the quantum uncertainty on the momentum of
$\Hb^+$ in the ground state of the harmonic Paul trap and to the
additional recoil associated with the photo-detachment of the extra
positron (see section \ref{sec:gbar}).

In this section we give an estimation of the uncertainty on the
arrival time associated with the initial vertical velocity spread
and show how it can be reduced by filtering out the fastest atoms.
Since slow antihydrogen atoms bounce on material surfaces thanks to
quantum reflection (see previous section), the filtering scheme used
in GRANIT with ultracold neutrons \cite{Nesvizhevsky2002nature}  can
also be applied in GBAR.

\subsection{Width of the arrival time distribution}

We consider a wave-packet falling in a linear gravitational
potential and want to determine the arrival time distribution on a
fixed horizontal plane, supposing there is no reflection from that
(ideal) detector. In this case classical and quantum calculations
give identical results, as can be seen by noticing that the Wigner
quasi-distribution function obeys the classical equations of motion
if the potential is at most quadratic. Therefore a given initial
phase-space distribution simply propagates along the classical
trajectories.

For a wave-packet initially centered at a height $H$ above the
detector, with zero mean velocity and uncorrelated vertical position
and velocity distributions of width $\Delta z$ and $\Delta v$
respectively, the spread of the arrival time distribution is
\begin{equation}
\frac{\Delta t}{t_H}= \sqrt{{\left(\frac{\Delta z}{2H}\right)}^2 +
{\left(\frac{\Delta v}{\sqrt{2 \gb H}}\right)}^2}~,
\label{eq:deltat}
\end{equation}
with $t_H=\sqrt{2 H/\gb}$ the classical free fall time. This
translates as a statistical uncertainty $\Delta \gb/\gb=2\Delta
t/\sqrt{N}t_H$ on the determination of $\gb$ after $N$ independent
measurements.

If the particle is initially in the ground state of a harmonic trap,
the distribution is  Gaussian and saturates the Heisenberg
inequality: $\Delta z\Delta v=\hbar/2m$. Then the  time uncertainty
is minimal for
\begin{equation} \Delta v=\Delta v_{opt} =
\sqrt{\frac{\hbar}{2m} \sqrt{\frac{\gb}{2 H}}}~.
\end{equation}
For $H=30$~cm and $\gb=g$ this evaluates to $\Delta v_{opt} \approx
3.6\times 10^{-4}$~m/s, and the relative uncertainty on the arrival
time is $2\times 10^{-4}$. However the current expected value for
GBAR is three orders of magnitude larger $\Delta v_0\approx
0.5$~m/s, which leads to a relative uncertainty of $0.2$.

The uncertainty in GBAR is largely dominated by the vertical
velocity dispersion. If the initial velocity dispersion can be
reduced from $\Delta v_0$ to $\Delta v$ by filtering out the hottest
atoms, the single-shot precision and the number of atoms are both
reduced by a factor $\Delta v/\Delta v_0$. Despite the loss in
statistics, this results in a net reduction of the statistical
uncertainty on $\gb$.

\subsection{Shaping of the vertical velocity distribution}

\begin{figure}
\centering
\includegraphics[width=0.80\textwidth]{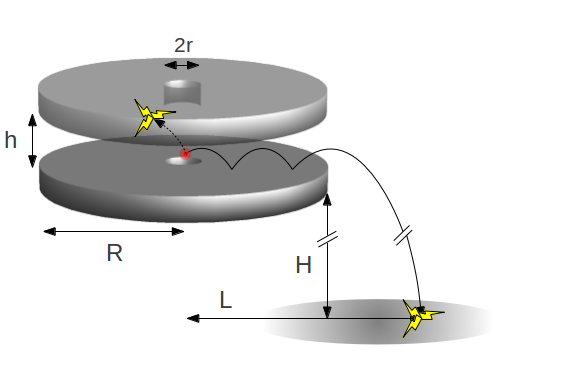}
\caption{\label{fig:shaper} Scheme of the proposed device to reduce the vertical velocity spread of the falling wavepacket (see text).}
\end{figure}

Our proposal \cite{Dufour2014} to realize this filtering is to let
the atoms pass through a horizontal slit between two  disks. The
bottom disk has a smooth top surface on which atoms reflect with
high probability whereas the top disk has a rough bottom surface
which effectively acts as an absorber for the atoms (see Fig.
\ref{fig:shaper}). Antihydrogen is initially trapped in the center
of the two disks (openings are made in the center of the disks to
allow operation of the trap). If its vertical velocity is high
enough to reach the rough surface, it is reflected non-specularly
and remains inside the device until it annihilates with high
probability. On the contrary, if it cannot reach the top disk the
atom will exit the device with high probability after bouncing on
the bottom mirror a few times. It then falls freely to a detector a
height $H$  below. Since the horizontal velocity is conserved, the
knowledge of the total time between photo-detachment and
annihilation and of the total horizontal $L$ distance traveled,
allows one to correct for the time spent inside the device before
the free fall.

If $h$ is the height of the slit, the velocity spread at the output
is $\Delta v\approx \sqrt{2 \gb h}$ and the proportion of atoms that
exit the device is $N/N_0 \approx \Delta v/\sqrt{2\pi}\Delta v_0$.
Using the shaping device therefore reduces the statistical
uncertainty on $\gb$ by a factor which scales as $h^{1/4}$.

Classically going to ever smaller slit heights leads to arbitrarily
good precision. For example for $h=1$~mm,  $\Delta v \approx
0.14$~m/s, $N/N_0\approx 5\%$ and the accuracy is improved by a
factor 2, whereas for  $h=50$~\micro m, $\Delta v \approx 0.03$~m/s,
$N/N_0\approx 1\%$ and the accuracy improved by a factor 4.

There are however two limits: the number of repetitions of the
experiment must be large enough that at least some atoms make it
through the filter, but more fundamentally, the wave function of the
atom must fit inside the slit. Indeed for slit sizes below 50 \micro
m the discrete spectrum of states bound by gravity must be taken
into account. For a slit size of 20 \micro m only the ground state
can travel through the guide, below that the transmission drops to
zero.

This fact has been used to demonstrate the existence of
gravitationally bound states for neutrons
\cite{Nesvizhevsky2002nature}. The next section explores the
possibilities of similar experiments with antihydrogen to further
increase the precision of equivalence principle tests on antimatter.

\section{Resonance spectroscopy of gravitational states of
antihydrogen near material surface}
\label{sec:gravstateshbar}

In this section we will study a motion of an $\Hb$ atom, localized
in a gravitational state near a horizontal plane mirror. The
existence of such states though counterintuitive is explained by the
phenomenon of quantum reflection of ultracold (anti-)atoms from a
steep attractive Casimir-Polder atom-surface potential. Such states
have similar properties with those discovered for neutrons
\cite{Nesvizhevsky2002nature,Nesvizhevsky2000,Nesvizhevsky2003,Nesvizhevsky2005,Voronin2006}.

To account for the interaction of $\Hb$ with a material  wall, the
gravitational quantum states \eqref{eq:En} receive a complex energy
shift $\varepsilon_0 \Delta$, with  $\Delta\simeq-i0.005$ for a
perfectly conducting wall \cite{Voronin2011}. All states therefore
acquire equal width, which is a function of a material surface
substance $\Gamma=2|\Delta|\varepsilon_0$. This width corresponds to
the lifetime of $0.1$~s in case of a perfectly conducting surface
and is twice longer for silica
\cite{Voronin2005pra,Dufour2013qrefl,Dufour2013porous} for instance.

The interest to study gravitational quasi-stationary states of $\Hb$
is due to their comparatively long life-time on one hand and easy
identification of certain state because of it's mesoscopic spatial
scale. This opens an interesting perspective to apply potentially
very precise resonance spectroscopy method to establish the
gravitational properties of anti-atoms. These methods are based on
inducing an observation of resonance transitions between
gravitational states. One of possible approach is to use an
alternating inhomogeneous magnetic field for such a purpose.

The interaction of a magnetic field with a ground state $H$ atom
moving through the field \cite{Lamb1952,Gorkov1968,Lozovik2004} is
dominated by the interaction of an average magnetic moment of the
atom \cite{Landau1965} in a given hyperfine state with the magnetic
field. We are going to focus on an alternating magnetic field with a
gradient in the vertical direction. This condition is needed for
coupling the field and the center of mass (c.m) $\Hb$ motion in the
gravitational field of the Earth. It allows one to induce resonant
transitions between quantum gravitational states of $\Hb$
\cite{Voronin2011}.

We will consider the magnetic field in the following form:
\begin{equation}\label{Magn1}
\vec{B}(z,x,t)=B_0 \vec{e}_z+ \beta \cos(\omega t) \left(z
\vec{e}_z-x \vec{e}_x \right)~.
\end{equation}
Here $B_0$ is the amplitude of a constant, vertically aligned,
component of magnetic field, $\beta$ is the value of magnetic field
gradient, $z$ is a distance measured in the vertical direction, $x$
is a distance measured in the horizontal direction, parallel to the
surface of a mirror. A time-varying magnetic field (\ref{Magn1}) is
accompanied with an electric field ($[\vec{\nabla} \vec{E}] =
-\frac{1}{c} \partial \vec{B} / \partial t$). However, for the
velocities of ultracold atoms, corresponding interaction terms are
small and thus will be omitted.

An inhomogeneous magnetic field couples the spin and the spatial
degrees of freedom. A $\Hb$ wave function is described in this case
using a four-component column (in a non-relativistic treatise) in
the spin space, each component being a function of the c.m.
coordinate $\vec{R}$, relative $\pb-\eb$ coordinate $\vec{\rho}$ and
time $t$. The corresponding Schr\"{o}dinger equation is:
\begin{equation} \label{Schr}
i \hbar\frac{\partial \Phi_{\alpha}(\vec{R},\vec{\rho},t)}{\partial
t} = \sum_{\alpha'}\left[ -\frac{\hbar^2}{2m}\Delta_R+Mgz+V_{CP}(z)
+ \widehat{H}_{in}+\widehat{H}_m \right]_{\alpha, \alpha'}
\Phi_{\alpha'}(\vec{R},\vec{\rho},t)~.
\end{equation}
A subscript $\alpha$ in this equation indicates one of four spin
states of the $\pb-\eb$ system. The meaning of the interaction terms
is the following. $V_{CP}(z)$ is an atom-mirror interaction
potential, which turns into the Casimir-Polder potential at an
asymptotic atom-mirror distance (see
\cite{Voronin2005pra,Voronin2005} and references therein).
$\widehat{H}_{in}$ is the Hamiltonian of the internal motion, which
includes the hyperfine interaction:
\begin{equation}\label{Hc}
\widehat{H}_{in} = -\frac{\hbar^2}{2
\mu}\Delta_\rho-e^2/\rho+\frac{\alpha_{HF}}{2}\left(\hat{F}^2-3/2\right)~.
\end{equation}
Here $\mu=m_1m_2/m$, $m_1$ is the antiproton mass, $m_2$ is the
positron mass, $m=m_1+m_2$, $\alpha_{HF}$ is the hyperfine constant,
$\hat{F}$ is the operator of the total spin of the antiproton and
the positron. We will treat only $\Hb$ atoms in a $1S$-state (below
we will show that the excitation of other states in the studied
process is improbable). The term
$\frac{\alpha_{HF}}{2}\left(\hat{F}^2-3/2\right)$ is a model
operator, which effectively accounts for the hyperfine interaction
and reproduces the hyperfine energy splitting correctly. The term
$\widehat{H}_m$ describes the field-magnetic moment interaction:
\begin{equation}\label{Hm}
\widehat{H}_m = -2\vec{B}(z,x,t)\left( \mu_{\eb}\hat{s}_{\eb}\times
\hat{I}_{\pb}+ \mu_{\pb}\hat{s}_{\pb}\times \hat{I}_{\eb}\right)~.
\end{equation}
Here $\mu_{\eb}$ and $\mu_{\pb}$ are magnetic moments of the
positron and the antiproton respectively, $\hat{s}_{\eb}$,
$\hat{s}_{\pb}$ is a spin operator, acting on spin variables of
positron (antiproton), $\hat{I}_{\eb}$, $\hat{I}_{\pb}$ is a
corresponding identity operator. As far as the field
$\vec{B}(z,x,t)$ changes in space and in time, this term couples the
spin and the c.m. motion.

We will assume that in typical conditions of a spectroscopy
experiment the $\Hb$ velocity component $v$ parallel to the mirror
surface (directed along $x$-axis) is of the order of a few $m/s$ and
is much larger than a typical vertical velocity in lowest
gravitational states (which is of the order of $cm/s$). We will
treat the motion in a frame moving with the velocity $v$ of the
$\Hb$ atom along the mirror surface. Thus we are going to consider
the x-component motion as a classical motion with a given velocity
$v$, and we will substitute a $x$-dependence by a $t$-dependence. We
will also assume that $B_0\gg \beta L$, where $L\sim 30$~cm is a
typical size of an experimental installation of interest. This
condition is needed for "freezing" the magnetic moment of an atom
along the vertical direction; it provides the maximum transition
probability.

We will be interested in the weak field case, such that the Zeeman
splitting is much smaller than the hyperfine level spacing $\mu_B
B_0\ll \alpha_{HF}$. The hierarchy of all mentioned above
interaction terms could be formulated as follows:
\begin{equation}\label{hierar}
m_2e^2/\hbar^2 \gg \alpha_{HF}\gg \mu_{\eb} |B_0 |\gg E_n~,
\end{equation}
and thus it justifies the use of the adiabatic expansion for solving
Eq. (\ref{Schr}); it is based on the fact that an internal state of
an $\Hb$ atom follows adiabatically the spatial and temporal
variations of an external magnetic field. Neglecting non-adiabatic
couplings, an equation system for the amplitude $C_n(t)$ of a
gravitational state $\psi_n(z)$ has the form:
\begin{equation}\label{Adiab}
i \hbar \frac{d C_{n}(t)}{dt} = \sum_{k} C_{k}(t) V_{n,k}(t)\exp
\left(-i\omega_{n k} t\right )~.
\end{equation}
The transition frequency $\omega_{n k}=(E_k-E_n)/\hbar$ is
determined by the gravitational energy level spacing. This fact is
used in the proposed approach to access the gravitational level
spacing by means of scanning the applied field frequency, as will be
explained in the following.

Within this formalism the role of the coupling potential $V(z,t)$ is
played by the energy of an atom in a fixed hyperfine state thought
of as a function of (slowly varying) distance $z$ and time $t$.
\begin{equation}
V_{n,k}(t)=\int_0^\infty \psi_n(z) \psi_{k}(z) E(t,z)dz~.
\end{equation}
Here $\psi_n(z)$ is the gravitational state wave function, which is
known in terms of the Airy function \cite{Voronin2011}.

The energy $E(z,t)$ is the eigenvalue of the internal and magnetic
interactions $\widehat{H}_{in}+\widehat{H}_m$, where the c.m.
coordinate $\vec{R}$ and time $t$ are treated as slow-changing
parameters. Corresponding expressions for the eigen-energies of a
$1S$ manifold are:
\begin{eqnarray}\label{Ea}
E_{a,c}&=& E_{1s} -
\frac{\alpha_{HF}}{4}\mp\frac{1}{2}\sqrt{\alpha_{HF}^2 +
|(\mu_B-\mu_{\pb})B(z,t)|^2},
\\ \label{Eb} E_{b,d}&=& E_{1s}+\frac{\alpha_{HF}}{4}\mp
\frac{1}{2}|(\mu_B+\mu_{\pb})B(z,t)|.
\end{eqnarray}
Subscripts $a,b,c,d$ are standard notations for hyperfine states of
a $1S$ manifold in a magnetic field. The presence of a constant
field $B_0$ produces the Zeeman splitting between states $b$ and
$d$. As far as the energy of states $b,d$ depends on magnetic field
linearly, while for states $a,c$ it depends quadratically, only
transition between $b,d$ states take place in case of a weak field.
In the following we will consider only transitions between
gravitational states in a $1S( b,d)$ manifold.

A qualitative behavior of the transition probability is given in the
Rabi formula, which can be deduced by means of neglecting the high
frequency terms compared to the resonance couplings of only two
states, initial $i$ and final $f$, in case the field frequency
$\omega$ is close to the transition frequency
$\omega_{if}=(E_f-E_i)/\hbar$:
\begin{equation}\label{Rabbi}
P = \frac{1}{2}\frac{(V_{if})^2}{(V_{if})^2 +
\hbar^2(\omega-\omega_{if})^2}
\sin^2\left(\frac{\sqrt{(V_{if})^2+\hbar^2(\omega-\omega_{if})^2}}{2\hbar}t\right)
\exp(-\Gamma t)~.
\end{equation}
The factor $1/2$ appears in front of the right-hand side of the
above expression due to the fact that only two $(b,d)$ of four
hyperfine states participate in the magnetically induced
transitions.

It is important that the transition frequencies $\omega_{if}$ do not
depend on the anti-atom-surface interaction up to the second order
in the splitting $\Delta$. This is a consequence of the already
mentioned fact that all energies of gravitational states acquire
equal shift due to the interaction with a material surface.

\begin{figure}
 \centering
\includegraphics[width=0.8\textwidth]{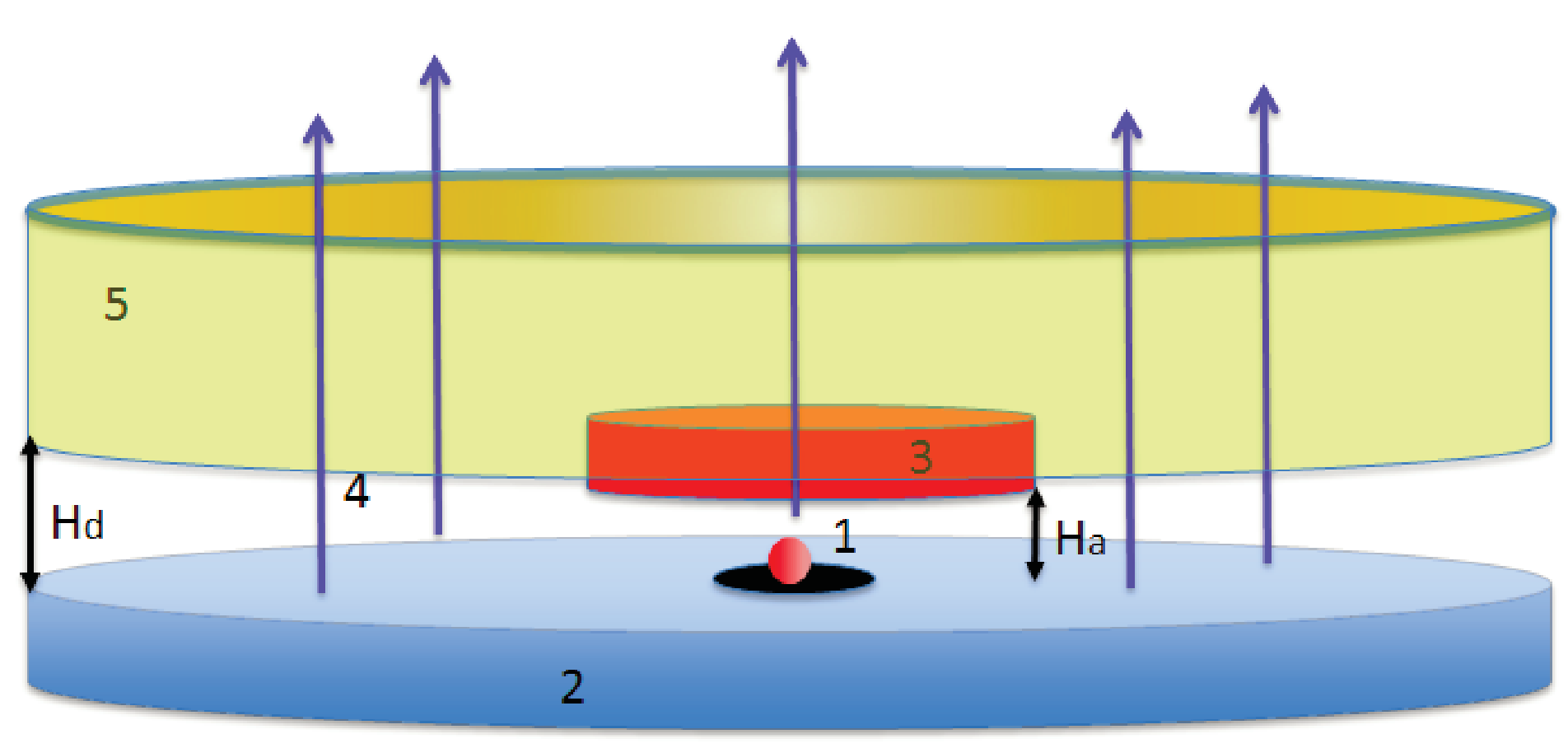}
\caption{A sketch of the principle scheme of an experiment on
magnetically induced resonant transitions between $\Hb$
gravitational states. 1 - a source of ultracold antihydrogen, 2 - a
mirror, 3 - an absorber, 4 - a magnetic field, 5 - a
detector.}\label{FigSketch}
\end{figure}

A resonant spectroscopy of $\Hb$ gravitational states could consist
of observing $\Hb$ atoms localized in the gravitational field above
a material surface at a certain height as a function of the applied
magnetic field frequency. A ``flow-through type'' experiment,
analogous to the one discussed for the spectroscopy of neutron
gravitational states \cite{Kreuz2009}, includes three main steps. A
sketch of a principle scheme of an experiment proposed in
\cite{Dufour2014} is shown in Fig. \ref{FigSketch}  (see also
section \ref{sec:shaper}).

First, an atom of $\Hb$ is shaped in a ground gravitational state.
This is achieved by means of passing $\Hb$ through a slit, formed by
a mirror and an absorber, which is placed above the mirror at a
given height $H_a$. The mirror and the absorber form a waveguide
with a state-dependent transmission \cite{Voronin2006}. The choice
of $H_a=H_1\simeq 13.6$~\micro m implies that only $\Hb$ atoms in
the ground gravitational state pass through the slit. Second, $\Hb$
atoms are affected by an alternating magnetic field (\ref{Magn1})
while they are moving parallel to the mirror. An excited
gravitational state is resonantly populated. Third, the number of
$\Hb$ atoms in an excited state is measured by means of counting the
annihilation events in a detector, which is placed at a height $H_d$
above the mirror. The value of $H_d$ is chosen to be larger than the
spatial size of the gravitational ground state and smaller than the
spatial size of the final state (\ref{eq:Hn}), $ H_1\ll H_d<H_f$, so
that the ground state atoms pass through, while atoms in the excited
state are detected.

\begin{figure}
 \centering
\includegraphics[width=0.8\textwidth]{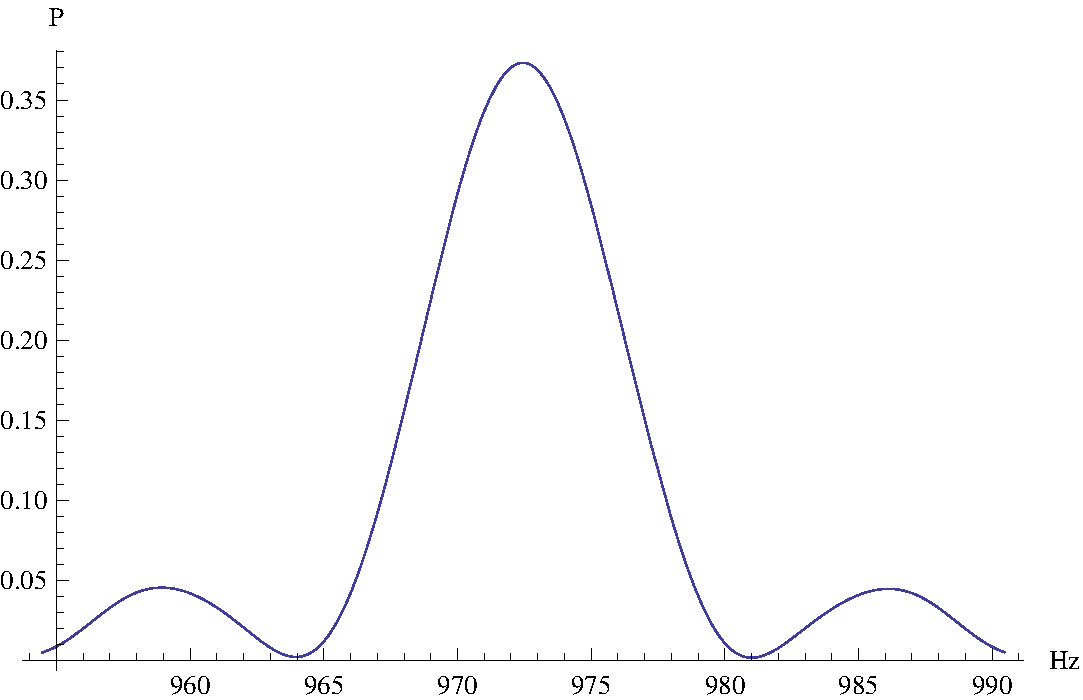}
\caption{The transition probability as a function of the magnetic
field frequency for the transition from the ground state to $6$-th
gravitational state.}\label{FigTrans}
\end{figure}

We present a simulation of the number of detected annihilation
events as a function of the field frequency in Fig. \ref{FigTrans}
for the transition from the ground to the $6$-th excited state,
based on a numerical solution of the equation system Eq.
(\ref{Adiab}). The corresponding resonance transition frequency is
$\omega=972.46$~Hz. The value of the field gradient, optimized to
obtain the maximum probability of $1\rightarrow 6$ transition during
the time of flight $t_{fl}=\tau=0.1$~s, turned to be equal
$\beta=27.2$~Gs/m, the corresponding guiding field value, which
guarantees the adiabaticity of the magnetic moment motion, is
$B_0=30$~Gs.

It follows from (\ref{eq:En}) that the $\Hb$ gravitational mass
could be deduced from the measured transition frequency
$\omega_{nk}$ as follows:
\begin{equation}
M=\sqrt{\frac{2m\hbar \omega_{nk}^3 }{g^2(\lambda_k-\lambda_n)^3}}~.
\end{equation}
Let us mention that $g$ in the above formula means the gravitational
field intensity near the Earth surface, a value which characterizes
properties of the field and is assumed to be known with a high
precision. At the same time all the information about gravitational
properties of $\Hb$ is included in the gravitational mass $M$.
Equality of the gravitational mass $M$ and the inertial mass $m$,
imposed by the Equivalence principle, results in the following
expression:
\begin{equation}
M=\frac{2\hbar \omega_{nk}^3 }{g^2(\lambda_k-\lambda_n)^3}~.
\end{equation}

Estimation of the accuracy of the above expression requires account
of different effects, including dynamical Stark shift of the
resonance line, non-adiabatic corrections to the transition
probability, interaction of alternating magnetic field with a
mirror, etc. The detailed study of different systematic effects is
under way. Assuming that the spectral line width is determined by
the lifetime $\tau\approx 0.1$~s of gravitational states, we
estimate that the gravitational mass $M$ can be deduced with the
relative accuracy $\epsilon_M\sim 10^{-3}$ for $100$ annihilation
events for the transition to the $6$-th state.

\section{Gravitational free fall of cold positronium}\label{sec:gravfallps}

\begin{figure}
\centering
\includegraphics[width=0.80\textwidth]{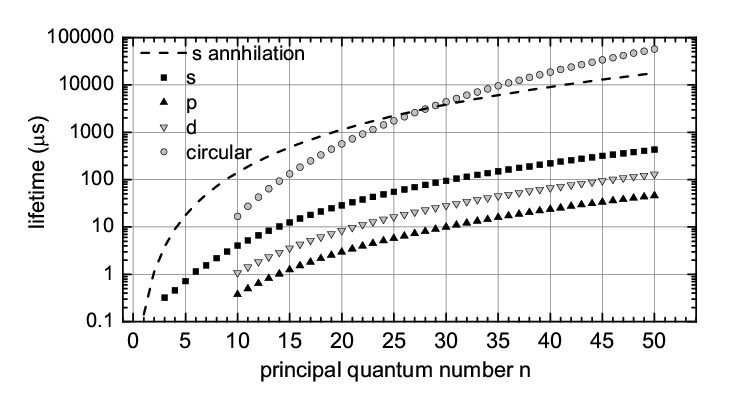}
\caption{\label{fig:lifetimesps} Radiative lifetimes of various Ps
states as a function of the principal quantum number $n$. The
lifetimes were calculated by summing the Einstein $A$ coefficients
of all electric-dipole-allowed decay channels from each Rydberg
state. For each $A$ coefficient the appropriate radial integrals
were determined using analytic expressions for the radial wave
functions in a pure Coulomb potential \cite{Bethe1957}. The dashed
line is the annihilation lifetime of $n$s states. After Ref.
\cite{Cassidy2014}. }
\end{figure}

Antihydrogen, muonium and positronium are the possible candidates
for gravity measurements on antimatter, with various pros and cons.
Antihydrogen and muonium \cite{Kirch2014} are extremely difficult to
produce, requiring large facilities (i.e., PSI, CERN), whereas
positronium is relatively easy to produce in smaller university
laboratories. However, Ps has an inconvenient propensity to
self-annihilate; the triplet ground state vacuum lifetime of only
142 ns, would seem to preclude using this system for a free fall
measurement. As has been pointed out by various authors, in
particular A. P. Mills, Jr., \cite{MillsJr.1989}, this is not the
case, since one need only excite Ps atoms into long-lived Rydberg
states to prevent self-annihilation. Indeed, for any Ps state with
$n>1$ the radiative lifetime is always less than the annihilation
lifetime\footnote{The only excited state for which this is not true
is the metastable 2s state.}. That is to say, for excited states the
overlap of the positron and electron wave functions is sufficiently
low that annihilation can be considered to be negligible (see Fig.
\ref{fig:lifetimesps}).

The radiative lifetimes of excited Ps states, shown in Fig.
\ref{fig:lifetimesps} , are almost twice those of the corresponding
states in hydrogen. For practical reasons the smallest Ps beam
deflections one can expect to observe will be 10's of micrometers or
more. Therefore, if Ps falls with the usual gravitational
acceleration, it would be necessary to produce states with lifetimes
of the order of a few ms to observe such deflections. As is evident
from Fig. \ref{fig:lifetimesps}, achieving such long radiative
lifetimes requires either exciting low $l$ Rydberg levels (i.e., s
or d) to extremely high principal quantum numbers, or going to lower
$n$ states (perhaps around $n$ = 30 or so) and then transferring the
atoms to circular states \bibnote{For a discussion of the properties
of circular states, and methods for producing them, see
\cite{Gallagher1994}} (or, if not true circular states, at least
states with higher angular momentum).

Aside from the creation of sufficiently long-lived Rydberg levels,
conducting a Ps free fall experiment will require solving many other
problems. In order to accomplish an experiment of the type first
outlined by Mills and Leventhal \cite{MillsJr.2002} it will be
necessary to produce a small (10-50 micron) ``point'' source of slow
positronium in a cryogenic environment. The resulting long-lived
Rydberg atoms will then have to be formed into a beam, perhaps by
electrostatic manipulation (focusing, and deceleration) via their
electric dipole moments \cite{Seiler2011}, and finally detected with
good spatial resolution (as a function of flight time) in order to
observe a deflection due to gravity. Possible methods to accomplish
some of these tasks are considered elsewhere
\cite{Cassidy2014,MillsJr.2002}.

\begin{figure}
\centering
\includegraphics[width=\textwidth]{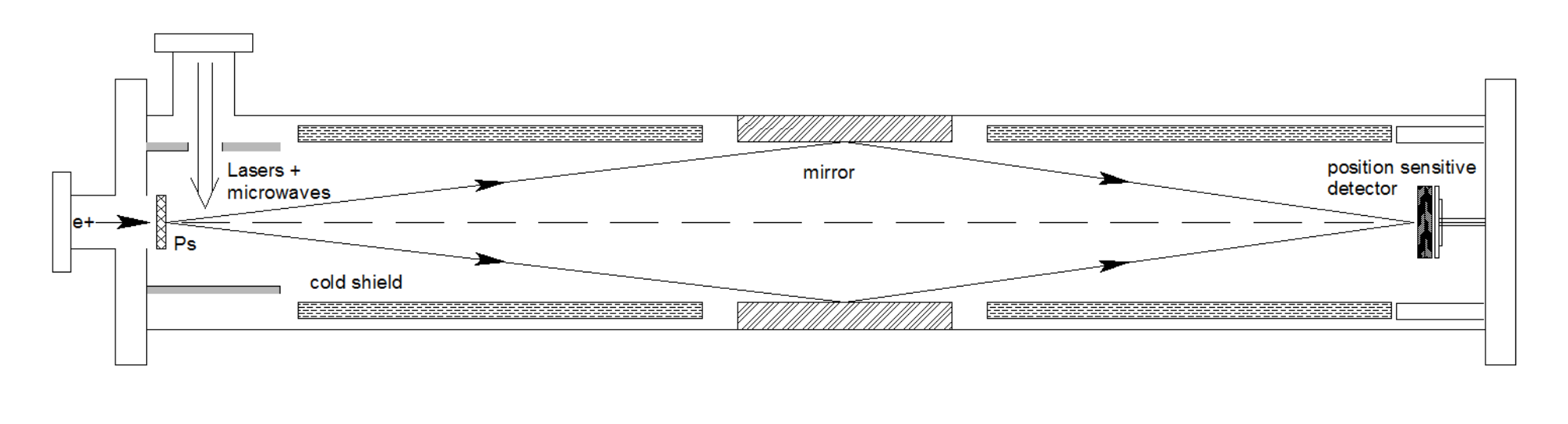}
\caption{\label{fig:expscheme} A schematic representation of a
Mills-Leventhal type of Ps free fall experiment. A real experiment
will undoubtedly be significantly different from this illustration,
which is intended only to highlight some of the different steps
involved. Of distinct practical concern will be the need to keep the
apparatus at low temperatures to mitigate effects of black body
radiation, as well as minimizing the Ps speed, which will determine
the length of the flight path, and hence the experiment. }
\end{figure}

The production of Ps Rydberg states with principle quantum numbers
around 30 can be accomplished using a two-step process
(1s$\to$2p$\to n$d), and has already been experimentally
demonstrated using broad-band ($\sim$ 100 GHz) lasers to accommodate
the large Doppler-broadened width of the transitions
\cite{Jones2014}. However, this methodology is not well suited to
the requirement that these atoms are subsequently transferred to
higher angular momentum states, and in order to achieve the required
state selectivity it may be necessary to use a different excitation
mechanism; i.e., a Doppler-free two-photon transition from the
ground state directly to a well-defined Rydberg Stark state
\bibnote{This has not yet been demonstrated for positronium; the
feasibility of doing so is considered in T. E. Wall, D. B. Cassidy
and S. D. Hogan, to be published.}.

As is well known, Rydberg atoms exhibit exaggerated properties
\cite{Gallagher1994} (see table \ref{tab:rydbergs}). In the present
case this is critical, since we seek to produce Ps atoms with very
long lifetimes, and also to take advantage of the large electric
dipole moments of Rydberg atoms to create and control an atomic
beam. However, insofar as we are compelled to make use of
electrically neutral systems to measure the weak gravitational force
acting on antimatter particles without extraneous electromagnetic
fields dominating their motion, excitation to states with very large
dipole moments brings us back to the original problem of extraneous
field effects. The situation is considerably less dire when dealing
with electric dipoles (and, to a much lesser extent, magnetic
dipoles) since in this case only field gradients give rise to
forces. Nevertheless, in an experiment designed to probe the weak
force of gravity with Rydberg atoms, forces due to stray fields must
be taken into account.

\begin{table}
  \centering
  \begin{tabular}{c|c|c|c|c}
& $n$-scaling & Ps & H & He \\ \hline\hline Binding energy (meV) &
$n^{-2}$ & -7.56 & -15.11 & -15.12\\ \hline State separation (meV) &
$n^{-3}$ & 0.48 & 0.96 & 0.96\\ \hline Orbital radius ($a_0$) &
$n^2$ & 2694 & 1347 & 1347 \\ \hline Radiative lifetime (\micro s) &
$n^3$ & 28.4 & 14.2 & 14.2\\ \hline Dipole moment/mass  ($e
a_0$/amu) & $n^2$ & 2.2 $\times 10^6$ & 1206 & 304
  \end{tabular}
\caption{\label{tab:rydbergs}The $n$-dependence of several
properties of Rydberg atoms, with examples shown for the 30d state
of Ps, H and He. The state separation is calculated for 30d $\to$
31d. The orbital radius is defined here as the expectation value
$\braket{r}  = \frac{1}{2} (3 n_{eff}^2 -l(l+1))$, where $n_{eff}$
includes the relevant quantum defect. The electric dipole
moment-to-mass ratios are calculated for the outermost state of the
$n$ = 30, $m$ = 2 Stark manifold. The radiative lifetime
$n$-dependence applies only to low $l$ states: for circular states
the scaling is closer to $n^5$ (see Fig. \ref{fig:lifetimesps}). }
\end{table}

\begin{figure}
\centering
\includegraphics[width=0.80\textwidth]{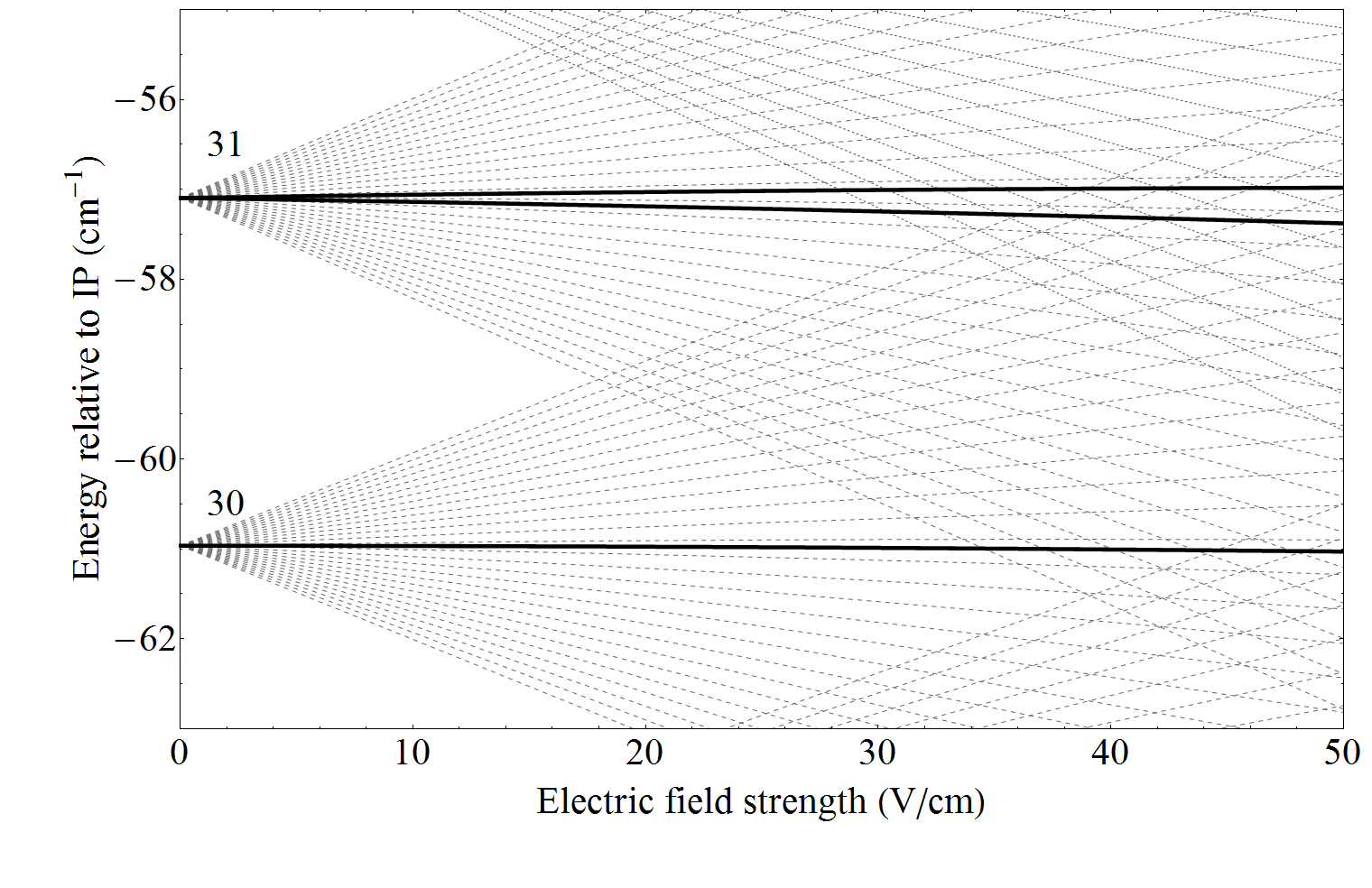}
\caption{\label{fig:starkstates} Stark states of $n$ = 30 and 31
states of Ps, with $m$ = 2 (grey dashed) and $m$ = 29 (black). In
the $n$ = 30 level the $m$ = 29 state is a circular state and
experiences no first-order Stark shift and only a very weak
second-order shift, as explained in the text.}
\end{figure}

When an atom is placed in an external electric field of strength $F$
and direction $z$, the field mixes the atom's angular momentum
states. To first order the state $\ket{n,l,m}$ is mixed with states
of adjacent $l$ but the same $n$ and $m$, \cite{Gallagher1988}. The
resulting Stark states repel each other, causing them to spread out
as the electric field strength is increased, as shown in Fig.
\ref{fig:starkstates}. Following the example of hydrogen, where the
first order Stark shift is analytically calculable, the
Schr\"odinger equation for an atom in an electric field can be
written in cylindrical coordinates, where the relevant quantum
numbers are $n$, $m$ and the parabolic quantum numbers $n_1$ and
$n_2$, which together satisfy the condition $n= n_1 + n_2 + |m| +1$.

The Stark states in a given $\ket{n,m}$ manifold are described by
the index $k=n_1-n_2$, where $k$ has values in the range from
$k_{min}=-(n-|m|-1)$ up to $k_{max}=n-|m|-1$ (with $\Delta k = 2$).
The first order Stark shift in Ps is given by $E_S=-\mu \times F$,
where the electric dipole moment has  magnitude $
|\mu|=\frac{3}{2}n|k| a_{\mathrm{Ps}}$ (where the Ps Bohr radius
$a_{\mathrm{Ps}}$ is (almost) twice that of hydrogen, i.e., $2
a_0$).  For a high-$n$ Rydberg state with low angular momentum, for
example the 30d state with $m = 2$, the value of the electric dipole
moment can be very large. The Stark state with $k_{max}=27$ has an
electric dipole moment of $2430 e a_0$. This large electric dipole
moment arises because within this $n$-state there are many
degenerate angular momentum states with the same value of $m$ that
are coupled by the electric field. While this is extremely useful
for atomic control \cite{Seiler2011,Vliegen2006,Hogan2008} it
presents a significant problem for gravity measurements, since the
electric field gradient experienced by a Ps atom in this state that
would result in a force equal to that of ``normal'' gravity ($\sim
2\times10^{-29}$~N ) is only $\sim 10^{-3}$~V/m$^2$. Although this
is by no means insignificant, it does compare favorably with the
$\sim 5\times 10^{-11}$~V/m electric field that would apply a
$g$-like force to a bare positron (or electron).

States with the maximum absolute values for the orbital and magnetic
quantum numbers for a given $n$, the so-called circular states,
experience no first-order Stark shift. For these states $m=|n|-1$,
meaning that there is only one Stark state associated with this
value of $m$, which has $k=0$, and thus, to first order, \emph{no
electric dipole moment} (see Fig. \ref{fig:starkstates}). The
explanation for this is that, within a given $n$-manifold, the
circular states have unique values of $m$, and thus are not coupled
to any other degenerate angular momentum states. In the classical
limit these states correspond to circular orbits, in which the
average $z$-position of the electron is zero, resulting in no
electric dipole moment, unlike the lower angular momentum Rydberg
states where, in the classical limit, the electronic orbit is highly
anisotropic, with the electron having a large average displacement
from the atomic core. Although there is no atomic core or nucleus in
the case of Ps, the wave function is nevertheless hydrogenic, and
the same arguments apply. The circular states do experience a
second-order Stark shift, from coupling of adjacent $n$-states,
however, this shift is extremely weak. It should, therefore, be
possible to produce Ps states with high angular momentum and
minimize the effects of stray fields while simultaneously extending
the lifetimes to useful levels. However, any manipulation techniques
that rely on large dipole moments will obviously have to be
performed after the optical excitation to the relevant $n$ states,
but before transferring the atoms to states with high angular
momentum.

Performing a gravity measurement on any system containing antimatter
is clearly very challenging, and many of the potential obstacles are
currently being investigated. The ability to produce controllable
beams of Ps atoms may also open the door to other types of
experiments, such as interferometry
\cite{Phillips1997,Oberthaler2002}, which could provide an
alternative route to an antimatter gravity measurement.

\section{Can we observe gravitational quantum states of positronium~? }
\label{sec:gravstatesps}

\begin{figure}
\centering
\includegraphics[width=\textwidth]{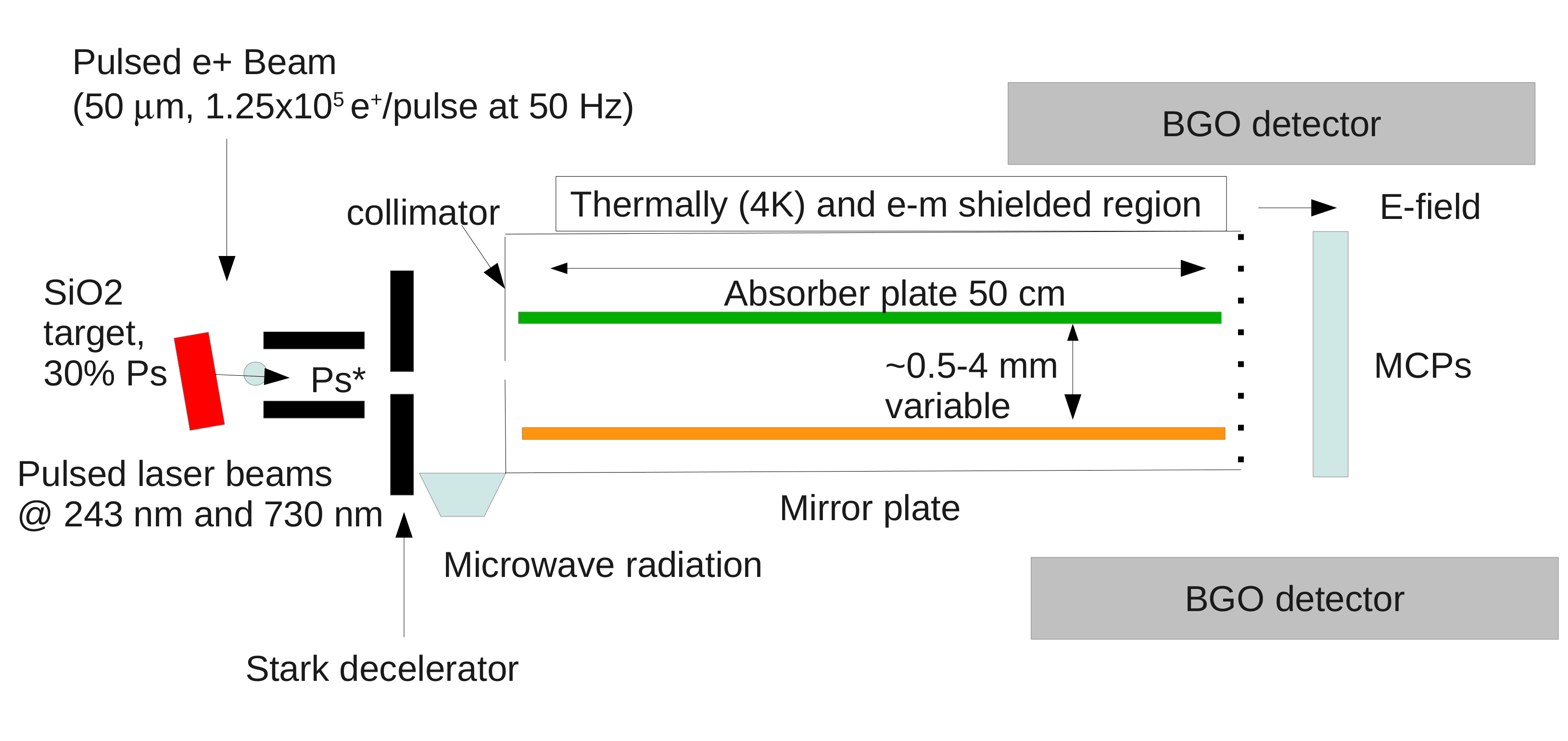}
\caption{\label{fig:SchemeQMBounce} Possible scheme for the
observation of the gravitational quantum states of positronium.}
\end{figure}

Positronium is about 1000 times lighter than a neutron or
antihydrogen.  Therefore the expected height of the gravitational
quantum state is 100 larger  corresponding to a macroscopic size of
$H_1 =1.3$~mm while the energy is 10 times smaller, $E_1=0.13$~peV
(see Eqs. \ref{eq:En}-\ref{eq:Hn}). The observation time to resolve
a quantum gravitational state can be estimated using the Heisenberg
uncertainty principle to be of the order of $\hbar/E_1 \simeq
4.5$~ms. This value is much larger than the long lived triplet
positronium lifetime in the ground state which is 142 ns (the Ps
singlet state only lives 125 ps and thus in the following we will
only consider the triplet state and refer to it as Ps).  Hence, as
for the case of a measurement of the gravitational free fall of Ps
described in the previous section, the Ps lifetime can be increased
by excitation to a higher level. A possible scheme to observe the Ps
gravitational quantum states could employ the flow-through technique
used for the first observation of this effect with neutrons (see
Fig. \ref{fig:SchemeQMBounce}).  Greater detail of the proposed
experimental set-up and technique are described in a dedicated
contribution to this workshop \bibnote{P. Crivelli, V.V.
Nesvizhevsky, and A.Yu. Voronin. Can we observe the gravitational
quantum states of Positronium? \emph{Advances in High Energy
Physics}, this issue, 2014.}. Here we describe the main idea.

Positronium is formed by implanting keV positrons from a
re-moderated pulsed slow positron beam in a positron-positronium
converter. To observe the quantum mechanical behavior of Ps in the
gravitational field its vertical velocity should be of the same
order of the gravitational energy levels and thus $v_y<0.15$~m/s.
Furthermore to resolve the quantum state the Ps atom has to interact
long enough with the slit and therefore it has to be laser excited
to a Rydberg state with $n>30$ and maximum $l$ quantum number (see
previous section). To keep a reasonable size of the experimental
setup (i.e a slit size of the order of 0.5~m) and minimize the
number of detectors the velocities in the horizontal plane should be
smaller than $v_{x,z}<100$~m/s. Similar to neutrons a collimator
could be used to select the velocity components $v_x,v_y$ of the
positronium distribution. However since no reliable thermal cold
source of positronium exists the velocity component perpendicular to
the surface $v_z$ has to be lowered by some other means. Relying on
the fact that atoms in Rydberg states have a large dipole moment
Stark deceleration can be used for this purpose.  This method has
been demonstrated for different atomic species (including hydrogen)
\cite{Hogan2008} and molecules \cite{Hogan2009}. Atoms in Rydberg
states have large dipole moments thus electric field gradients can
be used to manipulate them. The acceleration/deceleration $a$
imparted to the Rydberg atoms is given by:
\begin{equation}
a=76 \nabla F \frac{1}{m} n k ~,
\end{equation}
where $\nabla F$ is the gradient of the electric field in
V~cm$^{-2}$, $m$ the mass of the decelerated particles in atomic
units, $n$ and $k$ the Stark state quantum numbers. H atoms in
$n=25$ and an initial velocity of 700 m/s can be brought at rest in
3 mm  \cite{Hogan2008}.

For Ps being 1000 times lighter decelerations exceeding $10^9$~m/s
could be realized and therefore the vertical velocity of Ps emitted
from thin silica films with initial velocities of the order of
$10^5$~m/s \cite{Cassidy2010,Crivelli2010} could be reduced to below
100 m/s. Since one is interested only in decelerating the
distribution that is almost perpendicular to the surface of the Ps
target, one can expect for those atoms an efficiency close to 100
\%. This is confirmed by preliminary simulations \bibnote{Private
communication with Dr. C. Seiler.}. The collimator will be placed
after the deceleration stage and the microwave region where
circularly polarized radiation will spin up the Ps to the maximum
$l$ so that kicks to the momentum imparted to the atoms in the
vertical direction during these processes will be accounted for.

The fraction of atoms with $v_y<0.15, v_x,v_z<100$~m/s is estimated
to be of the order of $2 \times 10^{-9}$. After the collimator the
Ps will fly through the slit made of a mirror and the absorber. If
the distance between them is smaller than the first expected
gravitational state (i.e $<1$~mm) this will not be transparent and
therefore no signal will be detected above the expected background
in the detectors. If the width of the slit is increased to a value
lying between the first and the second gravitational state (i.e.
$<2$~mm) the Ps wave function can propagate and a signal is expected
to be detected via field-ionization and subsequent detection with
MCPs. This quantum jump would provide the unambiguous indication of
the observation of a quantum gravitational state of positronium.

As a mirror for Ps we propose to exploit a gradient of magnetic
field created using wires arranged parallel to each other with a
constant current to create a uniform gradient of the magnetic field.
Only the Ps triplet atoms with $m=0$ have a non-zero net magnetic
moment. For the $m=\pm1$ the electron and the positron magnetic
moments cancel and therefore those are insensitive to the magnetic
field. Therefore, only one third of the initial population will be
reflected. To equate the $E_y = 0.1$~peV a field of few mG at the
wire surface will be sufficient.

Because of the large spacial size of gravitational quantum states
and the very large characteristic length of the mirror needed to
form the gravitational states that is much larger than a
characteristic inter-wire distance,  we expect that the very weak
magnetic gradient will not perturb the gravitational states. The
strict theoretical analysis of this clearly mathematically defined
problem is ongoing. A matter mirror could also be considered. Due to
the large spacial size of the gravitational quantum state,  the
surface potential is expected to be very sharp and therefore result
in efficient quantum reflection (see section \ref{sec:qrefl}).In
both cases (magnetic or material mirror) we expect to have
effectively (quasi-classically) only a few collisions with the
surface. Nevertheless, the transitions rates due to quenching and
ionization caused by the electric or magnetic fields have to be
calculated. The absorber as for the neutrons is a rough surface on
which the impinging Ps will mix its velocity components and
therefore be lost.

With such a scheme assuming a mono-energetic slow positron beam flux
of $9\times 10^8$ e$^+$/s (this the highest intensity reported so
far reached at the FMR~II NEMOPUC source in Munich
\cite{Hugenschmidt2008}) an event rate of 0.8 events/day with a
background 0.05 events/day might be achievable with a realistic
extrapolation of current technologies. Possible losses due to
spurious effects like stray electric or magnetic fields or black
body radiation seems to be negligible but as for the case of a free
gravity fall further calculations and preliminary experiments should
be done to confirm this assumption and that all the required
efficiencies (e.g. Ps excitation in the $n=33$, $l=32$ state) can be
attained.

To note that the expected height of the gravitational state is
related to the gravitational mass $M$ by Eq. \eqref{eq:Hn}. This
means that for an uncertainty in the determination of $H_1$ of
$\delta H_1$ one can get an accuracy in the determination of $M$ at
the level of $\delta M/M= 3 \delta H_1/H_1\sqrt{N}$ where $N$ is the
number of detected signals. Assuming an uncertainty of $\delta H_1 =
0.1$~mm which is mainly determined by the finite source size the
value of $\delta M/M$ can be determined to 3\% in three months. This
is comparable to the accuracy that is aimed for antihydrogen
experiments at CERN
\cite{Kellerbauer2008,Vanderwerf2014,Zhmoginov2013}. Therefore,
observation of Ps gravitational quantum states offers a
complementary approach to test the effect of gravity on a pure
leptonic system. Most of the techniques required for such an
experiment are under development for the ongoing free gravity fall
experiment of Ps (see section \ref{sec:gravfallps}) and Rydberg Ps
deceleration experiments are being considered at ETH Zurich where
Prof. B. Brown's (Marquette University) buffer gas trap is being
commissioned. The advantage of using gravitational quantum states is
that unpredicted perturbations of the Ps atoms will not result in a
systematic effect for the experiment but will only affect the signal
rate. Therefore as for the case of antihydrogen this approach seems
promising to provide a much higher accuracy than a free fall
experiment.

\section*{Conclusion}

In this review, we have reported the progress of ongoing experiments
to measure gravitational free fall of antimatter. The GBAR
experiment will produce antihydrogen atoms in the ultracold regime
where quantum reflection from surfaces takes place. Quantum
reflection will allow the observation of gravitational quantum
states of antimatter that promise to lead to a very sensitive probe
of the effect of gravity on anti-atoms (2 orders of magnitude
improvement compared to the free fall experiments).

The techniques developed in experiments designed to produce a cold
beam of Ps for a free fall measurement will also eventually find
application in creating ultracold Ps atoms, as required for
observing gravitational quantum states. They will also enable a wide
variety of other experimental areas, such as precision spectroscopy.

Antimatter atoms in gravitational quantum states also provide a
unique opportunity to constrain experimentally extra short-range
forces between the mirror and the anti-atom with about the same
sensitivity as we  do for normal matter \cite{Antoniadis2011}.

\section*{Acknowledgements}

The authors wish to thank the GRANIT collaboration and the GBAR
collaboration (\href{http://gbar.in2p3.fr}{gbar.in2p3.fr}) for
providing excellent possibilities for discussions and exchange. In
addition to the involved institutes and laboratories, preparatory
work for GBAR has been funded by the ``Conseil G\'en\'eral de
l'Essonne'', by the ``Agence Nationale de la Recherche'' in France,
and by the LABEX P2IO.

P.C. acknowledges the support by the Swiss National Science
Foundation (grant PZ00P2\_132059) and ETH Zurich (grant
ETH-47-12-1). DBC and TEW gratefully acknowledge funding from the
EPSRC (EP/K028774/1), the Leverhulme Trust (RPG-2013-055) and the
ERC (CIG 630119).

\bibliographystyle{unsrt}
\bibliography{GRANIT}

 \end{document}